# Constraining the surface properties of Helene

C.J.A. Howett[1] and E. Royer[2]

1 - Southwest Research Institute, Colorado, USA.

2 – PSI, Tucson, AZ, USA

**Corresponding Author and their Contact Details:**

C.J.A. Howett

Email: howett@boulder.swri.edu

Telephone Number: +1 720 240 0120

Fax Number: +1 303-546-9687

Address:

1050 Walnut Street, Suite 300

Boulder, Colorado

80302

USA

**Abstract**

We analyze two sets of observations of Dione's co-orbital satellite Helene taken by Cassini's Composite Infrared Spectrometer (CIRS). The first observation was a CIRS FP3 (600 to 1100cm$^{-1}$, 9.1 to 16.7 µm) stare of Helene's trailing hemisphere, where two of the ten FP3 pixels were filled. The daytime surface temperatures derived from these observations were 83.3±0.9 K and 88.8±0.8 K at local times 223° to 288° and 180° to 238° respectively. When these temperatures were compared to a 1-D thermophysical model only albedos between 0.25 and 0.70 were able to fit the data, with a mean and standard deviation of 0.43±0.12. All thermal inertias tested between 1 and 2000 J m$^{-2}$ K$^{-1}$ s$^{-1/2}$ could fit the data (i.e. thermal inertia was not constrained). The second observation analyzed was a FP3 and FP4 (1100 to 1400cm$^{-1}$, 7.1 to 9.1 µm) scan of Helene's leading hemisphere. Temperatures between 77 and 89 K were observed with FP3, with a typical error between 5 and 10 K. The surface temperatures derived from FP4 were higher, between 98 and 106 K, but with much larger errors (between 10 and 30 K) and thus the FP3- and FP4-derived temperature largely agree within their uncertainty. Dione's disk-integrated bolometric Bond albedos have been found to be between 0.63±0.15 (Howett et al., 2010) and 0.44±0.13 (Howett et al., 2014). Thus Helene may be darker than Dione, which is the opposite of the trend found at shorter wavelengths (c.f. Hedman et al., 2020; Royer et al., 2020). However few conclusions can be drawn since the albedos of Dione and Helene agree within their uncertainty.

# 1 Introduction

Helene is a small satellite of Saturn, co-orbiting with Dione. It is located in Dione's Lagrangian point (L4), 6.26 Saturn-radii away from Saturn. Helene is non-spherical: 22.6 km by 19.5 km by 13.3 km (Thomas and Helfenstein, 2019). Due to its small size Helene is hard to study from the Earth (18.5 magnitude) but Cassini Imaging Science Subsystem (ISS) imaged it up to 40 m/pixel spatial resolution (Thomas et al., 2018). The highest-resolution imagery covered ~350° W eastwards to 220° W over approximately all latitudes (the sub-spacecraft latitude was 2° N). These images showed that Helene is covered in a 10-20m thick regolith that is being eroded at scarps (e.g. Figure 1) (Thomas et al., 2018). Umurhan et al. (2015) showed that non-steady state processes are probably the cause of downslope material movements, with rapid coverings that are followed by the downslope removal of additional loose material.

Studies show that Helene does not have the same spectroscopic properties as Dione (Verbiscer et al., 2007, 2018; Filacchione et al., 2012). Filacchione et al. (2012) used visible and infrared wavelengths as observed by Cassini's Visual Infrared Mapping Spectrometer (VIMS, 0.350-0.981 μm and 0.983-5.125 μm) also showed that Helene is bluer and hence richer in fresh water ice than Dione. Royer et al. (2020) used ultra-violet and visible data from Cassini's Ultraviolet Imaging Spectrograph Subsystem (UVIS), VIMS, and Imaging Science Subsystem (ISS) to show that Helene is more water-ice rich than Dione and/or less contaminated. Hedman et al. (2020) using Cassini ISS data (clear filter, with a central wavelength is 610 nm) showed that Helene is substantially brighter than Dione, which maybe due to its lower contamination by darker material.

Cassini's Composite Infrared Spectrometer (CIRS) had two targeted observations of Helene, neither of which has been previously analyzed. These data can provide insights into Helene's surface temperature and thermophysical properties. It is these data that we analyze here.

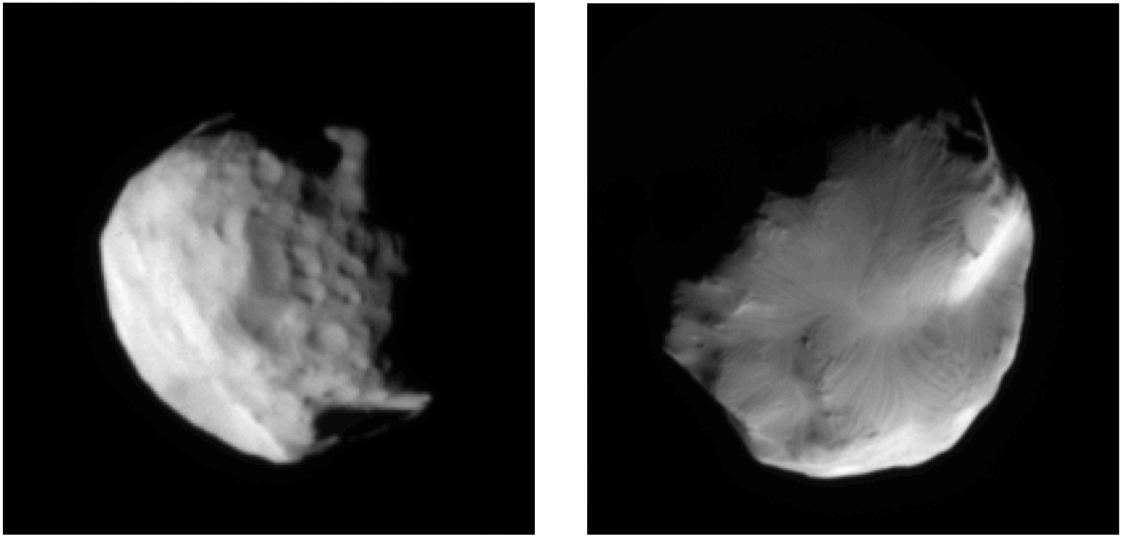

**Figure 1:** Images of Helene, obtained by the Cassini NAC during Revs 48 (left) and 144 (right) using its clear filters (in mode "CL1", "CL2"). Left: Image N1563643679, taken on 31st Jan 2011 at 10:21:25 UTC in Rev 144 (sub-spacecraft location is 94° W, 4° S). Right: Image N1675163339, taken on 20th July 2007 in Rev 48 at 16:53:26 UTC (sub-spacecraft location is 303° W, 2° S)

**2 Data**

CIRS is a Fourier Transform Spectrometer. The instrument has two interferometers that share a telescope and scan mechanism. Two thermopile detectors, known as focal plane 1 (or FP1), cover 10 to 600 cm$^{-1}$ (9.1 to 1000 μm) over a 3.9 mrad field of view. Mid-infrared emissions are detected by a Michelson interferometer and two linear focal plane arrays of 10 HgCdTe detectors, known as focal planes 3 and 4 (or FP3 and FP4). Both focal planes have a 0.273 mrad square field of view per detector, with FP3 sensitive from 600 to 1100cm$^{-1}$ (9.1 to 16.7 μm) and FP4 from 1100 to 1400cm$^{-1}$ (7.1 to 9.1 μm) (c.f Flasar et al., 2004). FP3 was an array of photoconductive detectors (PC), while FP4 is an array of photovoltaic (PV) ones. The layout of these arrays is given in Figure 2. We adopt this detector numbering system in analyzing the FP3 and FP4 observations.

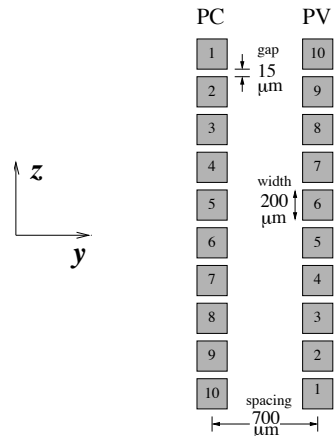

Figure 5: CIRS mid-infrared focal plane layout, showing detector spacing and numbering.

Figure 2 – CIRS' mid-infrared focal plane layout, showing the detector spacing and numbering (from Nixon et al., 2012).

CIRS made two targeted resolved observations of Helene during the lifetime of Cassini. These occurred on 20$^{th}$ July 2007 (Orbit, or Rev 48) and 31$^{st}$ January 2011 (Rev 144). Details of these observations are given in Table 1. The footprints of the FP3 (Rev 48), and FP3 and FP4 (Rev 144) are given in Figure 3. As the figure shows, in Rev 48 two of FP3 detectors (numbers 7 and 8) stared at Helene's leading hemisphere. These observations were made in centers mode, where the center five detectors of the FP3 focal plane are read out (as opposed to the usual pairs mode where odd numbered pixels are read out, followed by even ones). In Rev 144 two FP3 (detector numbers 4 and 5) and two FP4 detectors (detector numbers 6 and 7) scanned Helene's trailing hemisphere in pairs mode. These observations are certainly limited in nature, but allow us to constrain the surface temperature of Helene during these encounters.

We note that the latitudes and longitudes of the CIRS footprints in the nominal pointing files (POI files) assume Helene is spherical with a radius of 16 km (from the spice kernel file "cpck_rock_21Jan2011_merged.tpc"). We updated this kernel file to include the actual dimensions of Helene: 22.6±0.2, 19.6±0.3 and 13.3±0.2 km (Thomas and Helfenstein, 2019), and then used reconstructed Cassini pointing kernels to determine the actual footprints of the CIRS field of view as shown in Figure 3.

| Observation | Rev 48 FP3 | Rev 144 FP3 | Rev 144 FP4 |
|---|---|---|---|
| Date | 7/20/07 | 1/31/11 | 1/31/11 |
| Start Time (UTC) | 16:51:15 | 10:16:35 | 10:19:21 |
| End Time (UTC) | 16:59:13 | 10:17:29 | 10:20:30 |
| Start Scet | 1184950275 | 1296468995 | 1296469161 |
| Final Scet | 1184950753 | 1296469049 | 1296469230 |
| Minimum Sub-Spacecraft Longitude (°W) | 310.4 | 89 | 92.7 |
| Maximum Sub-Spacecraft Longitude (°W) | 314.7 | 90.3 | 94.1 |
| Sub-Spacecraft Latitude (°) | -3.6 | -7.7 | -7.6 to -7.7 |
| Minimum Sub-Solar Longitude (°) | 5.6 | 180.8 | 181 |
| Maximum Sub-Solar Longitude (°) | 6.2 | 180.9 | 181.1 |
| Sub-Solar Latitude (°) | -11.5 | 8.2 | 8.2 |
| Minimum Range (km) | 38,635 | 27,661 | 27,679 |
| Maximum Range (km) | 39,006 | 27,662 | 27,701 |

Table 1 – Details of the Helene observations analyzed here. "Scet" describes the time on the spacecraft clock, and range describes the distance between Cassini and the surface of Helene. The range is the distance between the spacecraft and the body center.

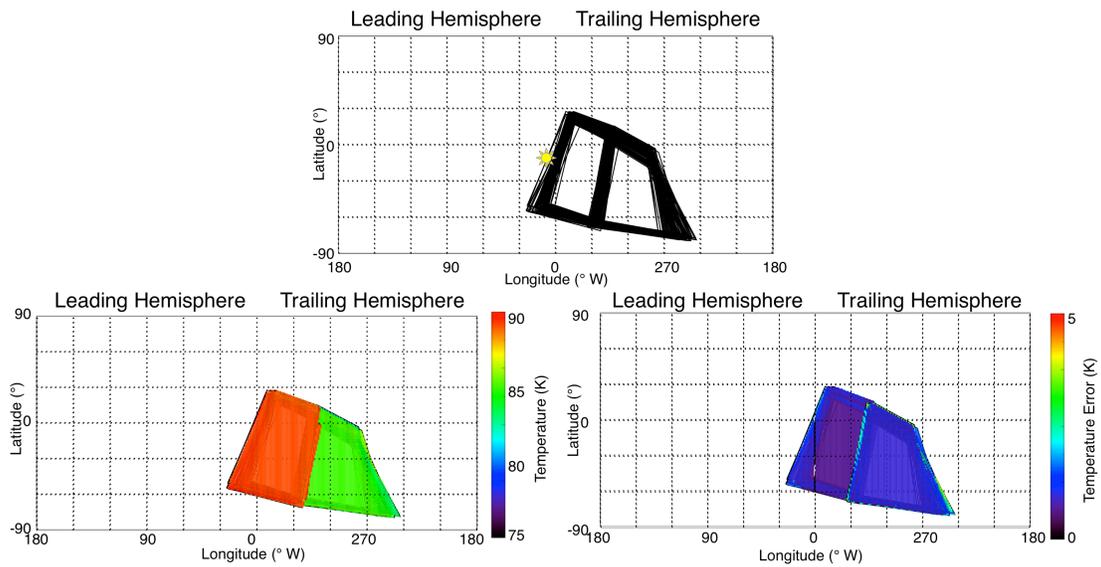

(a) CIRS FP3 observations of Helene during Rev 48.

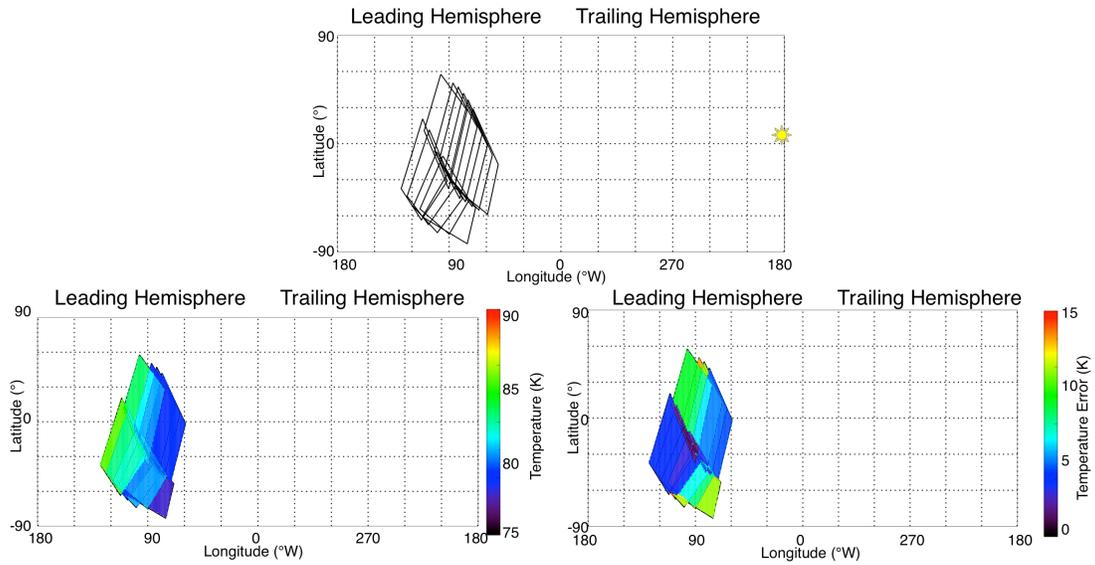

(b) CIRS FP3 observations of Helene during Rev 144.

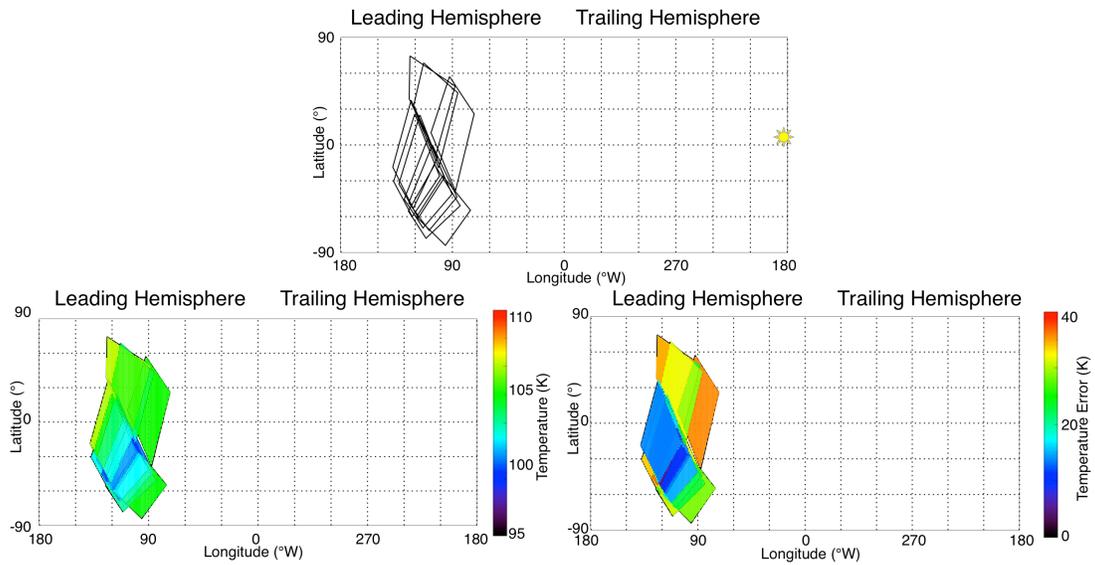

(c) CIRS FP4 observations of Helene during Rev 144.

**Figure 3**. The footprints, derived surface temperature and their corresponding error for all resolved CIRS observations of Helene. Top: the footprint of all resolved CIRS

resolved observations of Helene. The location of the sub-solar point is indicated by the Sun-shaped icon. Bottom Left: Helene's surface temperature derived from the CIRS observations, when fields of view overlap the best fitting blackbody temperature is found to the average radiance. Bottom Right: The error on the derived surface temperatures, see text for how they were derived. Note that the longitude convention we use here is different to what is normally used for Saturnian icy satellites (usually plotted with 180° W at the center). This was done so that the Rev 48 fields of view were not split across the 0° W boundary.

## 3. Data Analysis

### 3.1 Surface temperature derivation

#### 3.1.1 Rev 48

For each detector the spectra it took during the Rev 48 stare observation can be co-added together to reduce the noise, since they observe the same terrain under approximately identical geometry. The resulting spectra are shown in Figure 4. As the figure shows the spectra get noisier towards higher wavenumbers, which is expected in the CIRS spectra as the blackbody temperature emission decreases at higher wavenumbers. The best-fitting blackbody spectrum is then found for this mean radiance. This fit was determined by fitting a blackbody curve to the observed emission using the downhill simplex method (c.f. Nelder and Mead, 1965) in IDL's amoeba algorithm, on the assumption that the surface emits as a blackbody. The noise on the derived surface temperatures is derived using a two-step Monte Carlo technique: first a synthetic noise with a comparable magnitude to the observed noise is created and added to the previously determined best fitting blackbody curve. Then this spectrum is fitted by a blackbody emission spectrum.

This process is repeated numerous times, and the temperature error estimate is given by the standard deviation of the temperatures whose blackbody emission spectra are best able to fit the created spectra. The best fitting blackbody temperatures for FP3 detectors 7 and 8 (with their 1σ error) are 83.3±0.9 and 88.8±0.8 K respectively, which are compared to their mean spectra in Figure 4.

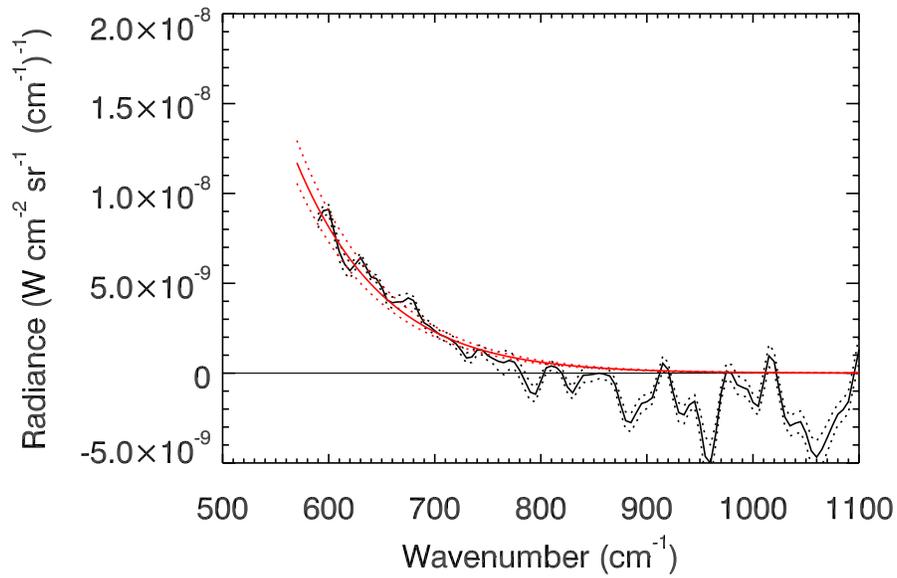

(a) Mean radiance and blackbody temperature fit (83.3±0.9 K) for detector 7

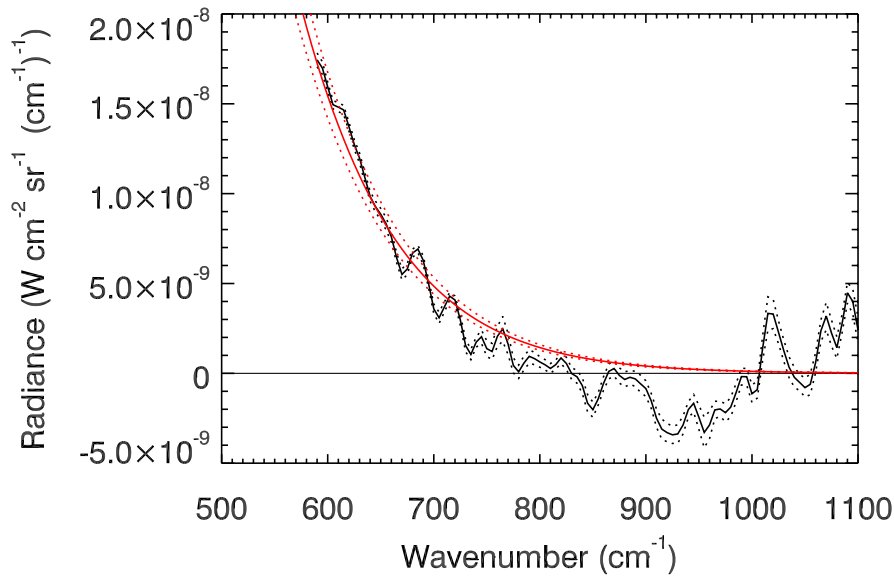

(b) Mean radiance and blackbody temperature fit (88.8±0.8 K) for detector 8

**Figure 4** – The mean FP3 CIRS radiance (black) with the standard error of the mean also shown (black dotted line). The mean blackbody fit to it also shown (red) with its error (red dashed line). The spectral resolution of CIRS in this range is 5 cm$^{-1}$, meaning there are 112 points in each of these spectra.

3.1.2 Rev 144

We use the same technique as previously described to derive the FP3 and FP4 surface temperatures from the Rev 144 observations. However, there is one significant difference: unlike the Rev 48 observations, those made in Rev 144 include both FP3 and FP4 observations, and it is a scan instead of a stare. Since these scans cover a variety of terrains, local times and geometries they cannot be simply coadded to increase the signal to noise like those in Rev 48. Thus, individual spectra have to be analyzed, which have a much lower signal to noise than the Rev 48 coadded spectrum, and thus the uncertainty of the derived surface temperatures are higher. As expected, and shown in Figure 3, FP3 has better sensitivity to the surface temperatures of Helene than FP4. This is because the wavenumber coverage of FP3 is closer to the peak of Helene's blackbody curve than FP4. The errors typical on the blackbody temperature derived by FP3 are between 5 and 10 K, compared to 10 to 30 K for FP4.

3.2 Albedo and Thermal Inertia Derivation

We only consider the temperature results from the Rev 48 observations, since those from Rev 144 have uncertainties too great to provide an adequate constraint. To derive Helene's thermophysical properties we follow a similar technique to the one we used to produce bolometric Bond albedo and thermal inertia maps of Rhea, Dione, Tethys and Mimas (Howett et al., 2014, 2019, 2020). Model diurnal temperatures were pre-calculated for each encounter (i.e. for a specific target rotation speed, latitude, local time, sub-solar latitude and heliocentric distance) by a simple diurnal 1-D thermal model (*thermprojrs*, c.f. Spencer, 1989). So while the model is a diurnal model the seasonal variation in heliocentric distance and sub-solar latitude are accounted for. The model assumes a unit emissivity and does not include heating from Saturn or reflected sunlight from Saturn, which are assumed to be negligible.

The model calculates, in one-dimension, the heat flow conducted to and away from the surface to determine the temperature as a function of depth and time of day. The upper boundary is set so that the thermal radiation and incident solar radiation are balanced with the heat conducted to and from the surface and the change in the heat content of the surface layer. The lower boundary is set to a depth at which there is negligible temperature change with the diurnal temperature cycle.

We note that our thermal model assumes Helene is spherical. Helene's elongated shape is accounted for since we have derived accurate planetographic longitude and latitudes from spice (which are by definition perpendicular to the surface of the spheroid) and Helene is not concave. Some errors maybe introduced by this assumption, for example the effect of reradiated sunlight from surfaces in small concaved regions will not be accounted for. However, we expect this effect to be minimal.

The model was run for a range of thermal inertias and bolometric Bond albedos: thermal inertias between 1 and 2000 J m$^{-2}$ K$^{-1}$ s$^{-1/2}$ (these units are henceforth referred to as MKS) and bolometric Bond albedos between 0.1 and 1.0, as shown in Figure 5. These values were selected as they probe the full range of expected bolometric Bond albedos, and thermal inertias between very porous ice (1 MKS) and bulk crystalline ice (2000 MKS, Ferrari and Lucas, 2016).

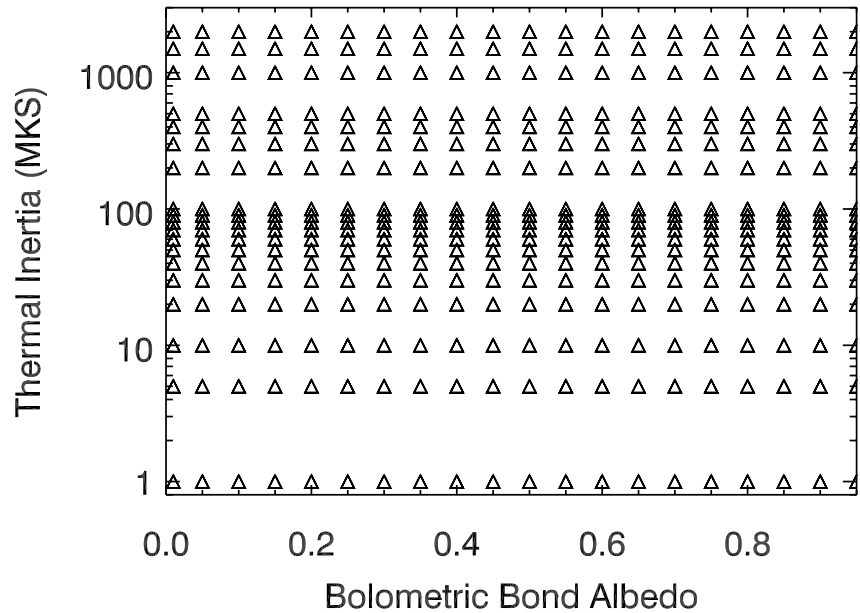

**Figure 5** - The range of bolometric Bond albedo and thermal inertia values modeled.

For each CIRS observation, the modeled surface temperatures were then compared to those determined from the data. To do this we determine the surface temperature predicted by a given thermal inertia and bolometric Bond albedo combination. The exact location of the field of view changes slightly with each of the 101 observations made (due to changing spacecraft altitude) so each of the observations is first considered separately. First the predicted surface temperatures at each of the four corners of the FP3 focal plane are calculated. Then, since the response of FP3 is flat (Nixon et al., 2009), the average surface temperature of the stare observation for a given albedo and thermal inertia combination is assumed to be the mean of these surface temperatures. This mean surface temperature is then converted to a blackbody spectrum, and the difference between that spectrum and the co-added CIRS spectrum (shown in Figure 4) is used to determine the reduced $\chi^2$ for the given albedo and thermal inertia combination. This process is then repeated for all albedo and thermal inertia combinations, and the results are shown in Figure 6. This result was shown to be insensitive to the high wavenumber cutoff (i.e. if it was reduced from 1100 cm$^{-1}$ to 900 cm$^{-1}$). All thermal inertias and albedo combinations that produce a reduced $\chi^2$ of less than 1 when compared to the mean spectrum observed (Figure 4) are assumed to be feasible. The results from the two detectors are not combined, because the local times they cover (and hence the temperatures predicted) vary significantly between the two observed scenes (Figure 4). As Figure 6 shows neither detector is able to constrain thermal inertia, but detector 7 provides an upper limit of Helene's albedo of 0.75 and detector 8 constrains it to between 0.25 and 0.70.

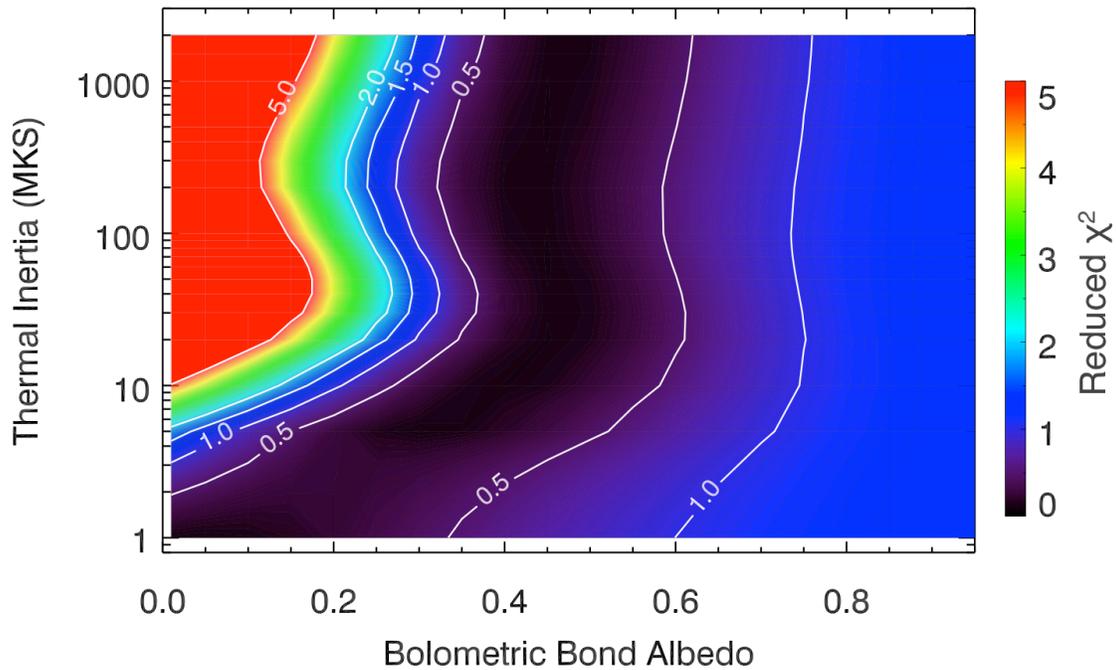

(a) FP3 detector 7

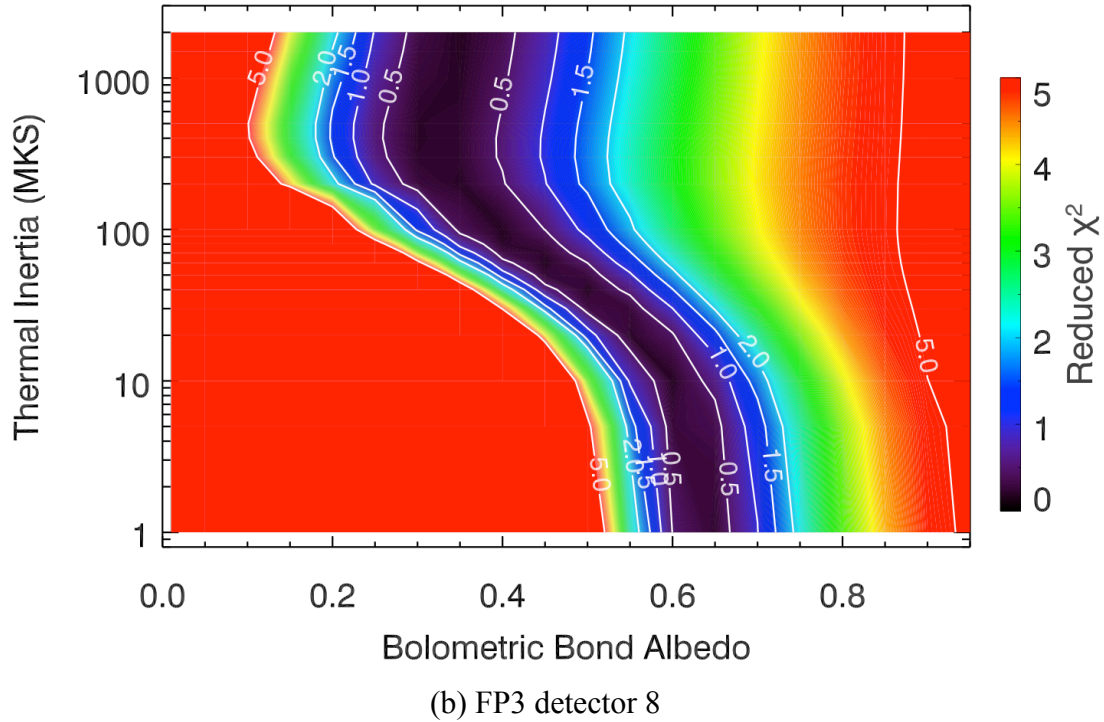

(b) FP3 detector 8

**Figure 6** - The reduced $\chi^2$ for the Rev 48 stare observations.

## 4. Discussion

The stare results from Rev 48 provide a high signal-to-noise observation, with a temperature error of <1 K. The two observations probe different local times: detector 7 (84±1.8 K) covers between 223° and 288° (i.e. early evening), and detector 8 (90.0±1.7 K) covers between 180° and 238° (i.e. afternoon cooling). So these temperatures follow expected warming patterns: the region closer to the sub-solar point (detector 8) is warmer than the one further from it. Such daytime temperatures are consistent with those seen on other saturnian icy satellites (Howett et al., 2010).

Analysis of the temperatures derived from detector 8 data constrain Helene's bolometric Bond albedo to be between 0.25 and 0.70, with a mean and standard deviation of

0.43±0.12. As noted above the thermal inertia could not be constrained and anything between a highly porous surface (low thermal inertia) and pure ice (2000 MKS surface) can fit the data. The reason detector 8 provides a better albedo constraint than that of detector 7 is not a factor of temperature uncertainty, their predicted temperatures and errors are very similar. Rather, the local times covered by the two detectors produce different sensitivities to the albedo and thermal inertia ranges probed, which is shown more clearly in Figure 7. At the local times covered by detector 7 the surfaces are cooling and all pass through ~ a single point around 240°, which is where this observation occurs. This makes it difficult to distinguish between different thermophysical property combinations. However, the local times covered by detector 8 are during the late afternoon, where different albedo models produce different ranges of warming on the surface, and thus its possible to distinguish between some of the thermophysical property combinations. Thus, detector 8 is better able to distinguish between these albedo models, not because of any additional temperature sensitivity but because the local times it covers are more diagnostic than those of detector 7.

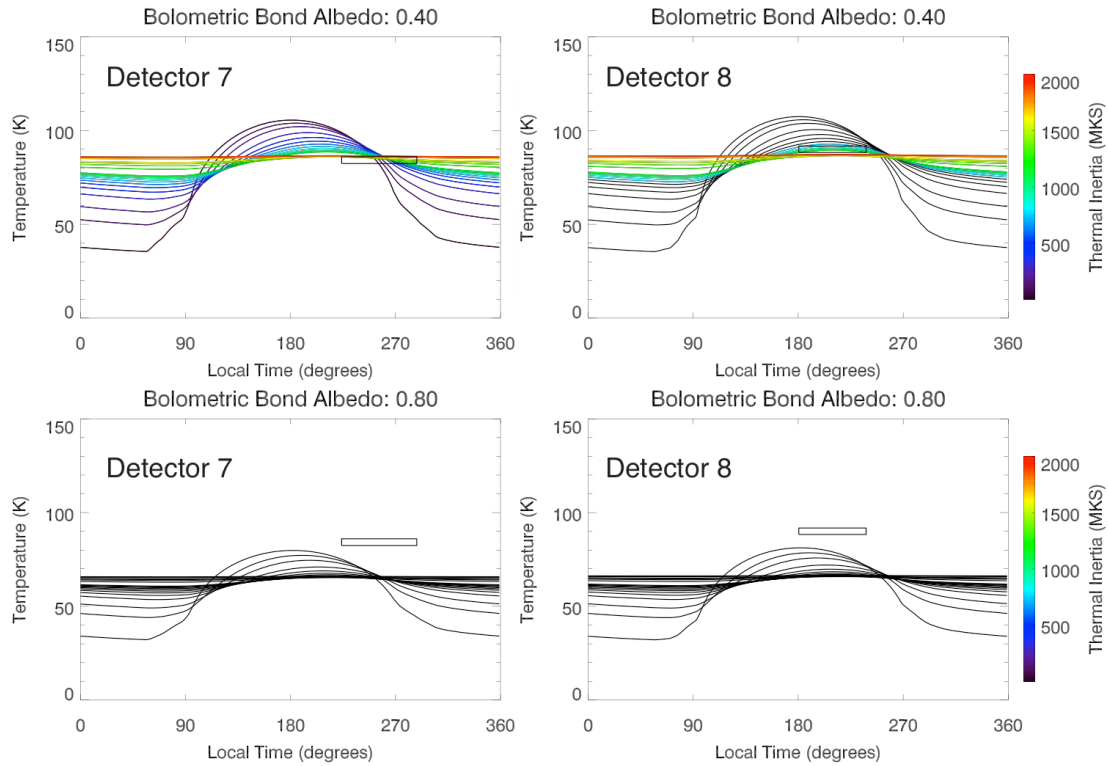

Figure 7- Predicted diurnal curves for two albedos for both FP3 observations. The exact observation shown is arbitrarily chosen (the 50$^{th}$ observation of both detectors) and taken at: 16:53:51 UTC (scet 1184950431) for detector 7, and 16:54:06 UTC (scet 1184950446) for detector 8). They are representative of all observations taken. In each figure all of the modeled thermal inertia runs are shown, while those in black do not fit the data (i.e. reduced $\chi^2>1$), while those shown in color do reproduce the observed temperatures (i.e. have a $\chi^2 \leq 1$). The colors denote the modeled thermal inertia, the modeled bolometric Bond albedo is given in the title of each of the sub figures. The black rectangle indicates the local times spanned by the CIRS observation and its temperature with uncertainty.

Howett et al. (2014) derived maps of both albedo and thermal inertia for Dione. These maps are given in Figure 8, which shows the thermal inertia varies from 8 to 11 MKS, and Dione has albedos ranging from 0.2 to 0.6. Since we were unable to constrain thermal inertia we cannot comment on how the thermal inertias of Helene and Dione compare, only to state that similar low thermal inertias to Dione could be present on Helene too (and all icy saturnian icy satellites, c.f. Howett et al., 2010). Howett et al. (2014) showed that the albedo of Dione's trailing (leading) hemisphere is 0.39 ± 0.13 (0.49 ± 0.11), with a disk-integrated value of 0.44 ± 0.13. While, Howett et al. (2010) used a more limited CIRS data set to derive an albedo for Dione of and 0.63 ± 0.15. Pitman et al. (2010) used data from Cassini's Visual and Infrared Mapping Spectrometer (VIMS) to derive an albedo of 0.52 ± 0.08. The albedo of Helene's trailing hemisphere derived here (0.25 to 0.75) covers such a wide range it is difficult to say conclusively whether it is brighter than Dione. However, there is a general trend of our derived albedo for Helene being less than Dione's, which is the opposite of other studies. For example, Hedman et al. (2020) and Royer et al. (2020) both showed that Helene was brighter than Dione. Hedman determined a brightness coefficient (done to account for the elongation of the small moons in his study) for Helene and Dione using Cassini's Imaging Science Subsystem (ISS) data. This coefficient is not an albedo, but does provide a measure of absolute brightness of the two objects, and showed Helene to be brighter than Dione (i.e. the reverse of the trend we find here). The reason why Helene is brighter than Dione is unclear, Hedman et al. (2020) suggest it maybe due to a localized increase in particle flux at Helene. There is not an obvious candidate for this, and Hedman suggest either a previously-unknown asymmetry in the E-ring particle flux or a transient phenomenon due to a recent event like an impact. Royer et al. (2020) analyzed data taken by Cassini's Ultraviolet imaging spectrograph (UVIS), ISS and the Visual and Infrared Mapping Spectrometer (VIMS) of Dione and Helene. Their results showed that the phase curve of Helene at 610 nm gives a single scattering albedo of 0.99, which indicates the surface must have extremely low contamination. As stated before, Helene's lower contamination may explain why Hedman et al. (2020) found it brighter than Dione. However, again we note that due to the large uncertainties of our results it is not clear whether they do indeed contradict those of Hedman et al. (2020) and Royer et al. (2020).

Making an assumption about Helene's thermal inertia can be used to better constrain its albedo. For example, if we assume that Helene has the same thermal inertia as Dione (10 MKS, Howett et al., 2014) then its albedo must be 0.60±0.05 for the region observed by FP3 detector 8, and 0.45±0.25 for the region observed by FP3 detector 7. While such an assumption about its thermal inertia is reasonable, since low thermal inertias are seen on all icy satellites throughout the Saturn-system (Howett et al., 2010), these results must be treated with caution (i.e. the result is only as correct as the assumption). We note that this result for detector 7 does not provide a better constraint than the one previously derived by detector 8 (0.43±0.12), and thus is of limited use.

Figure 3 shows the temperature maps derived from both FP3 and FP4 scans for Rev 144, which have overlapping coverage. Temperatures between 77 and 89 K were observed with FP3, with a typical error between 5 and 10 K. While the surface temperatures derived from FP4 were higher (between 98 and 106 K) with errors between 10 and 30 K.

It is unclear what is driving these differences, but one possible explanation is that the focal planes are probing slightly different surface depths (with FP4's shorter wavelength probing less deep than FP3) and thus are observing slightly different surface temperatures.

Since the Rev 48 data are from a scan they could not be coadded, resulting in significantly noisier data than those from Rev 48. The FP3 and FP4 regions overlap in some locations, and the results agree within their (large) uncertainties. The temperatures generally follow the trend of warming towards the sub-solar point. The main exception to this is in FP4, which appears to display a warmer region centered at ~110 °W/40 °S. However, the uncertainty in the results is greater than the magnitude of the warming, and nothing similar is seen in the FP3 map (although we note the FP3 map does not extend as far north). Thus, we cannot conclude conclusively that this region is thermally anomalous.

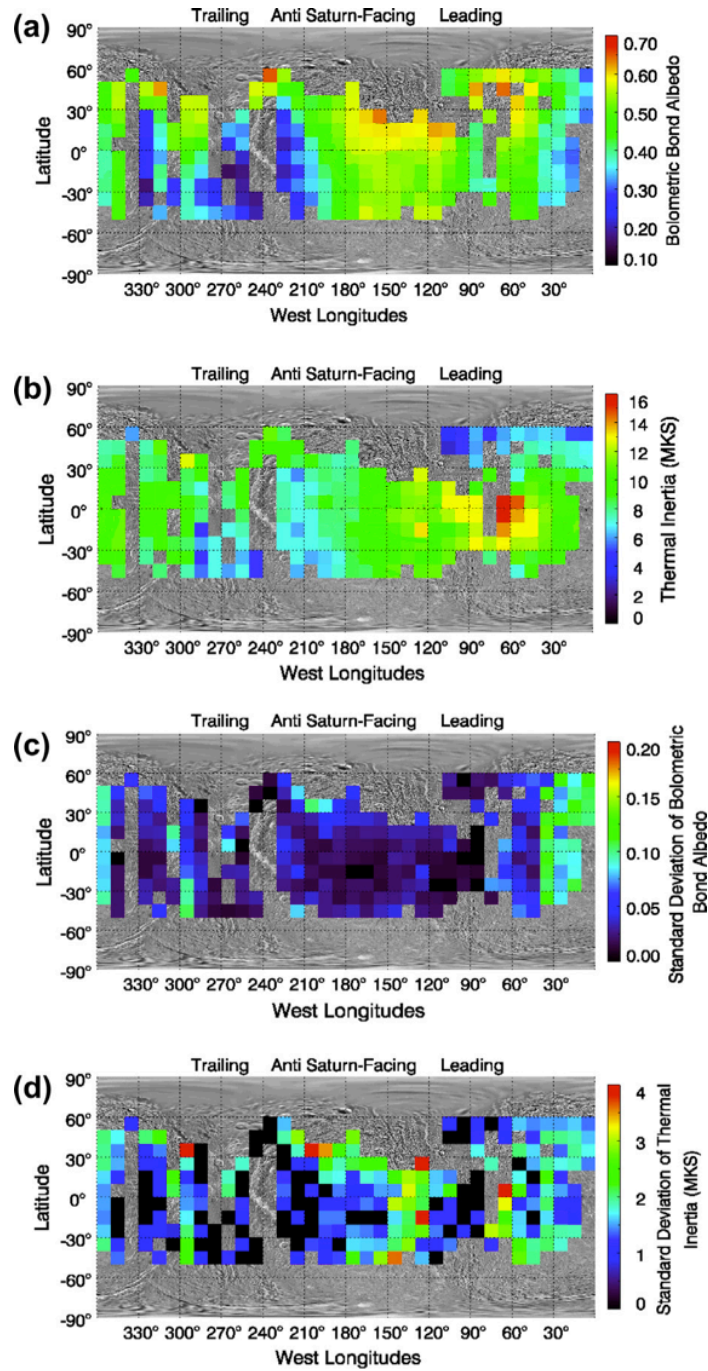

Figure 8: Maps of derived thermophysical properties for Dione from Howett et al. (2014). (a) Bolometric Bond albedo map of Dione. (b) Thermal inertia map of Dione. (c) The standard deviation of the bolometric Bond albedo values given in (a). (d) The standard deviation of the thermal inertia values given in (b).

**Conclusions**

The Rev 48 FP3 stare observation of Helene proved to be the most useful in analyzing Helene's thermophysical surface properties, largely because it enabled the spectra to be coadded, which significantly reduced the error of the derived surface temperatures. The thermal inertia of the surface could not be constrained, but limits on Helene's bolometric Bond albedo could constrained to be between 0.25 and 0.70. The mean and standard deviation of the model albedos able to fit the data is 0.43±0.12. This albedo is darker than that of Dione, which contradicts other work in the literature. However, the uncertainty on our result is too large to draw any strong conclusions from. Scans of Helene by FP3 and FP4 were able to produce temperature maps, but the uncertainty on the derived surface temperatures were too large to make further analysis possible.

**Acknowledgements**

We thank the Cassini Data Analysis Program for their support of this work (NNH16ZDA001N-CDAP), and John Spencer for his helpful guidance.

**References**

Ferrari, C., and A. Lucas, Low thermal inertias of icy planetary surfaces, A&A 588, A133, 2016.

Filacchione G. et al., Saturn's icy satellites and rings investigated by Cassini-VIMS: III — Radial compositional variability. Icarus, 220, 1064–1096, 2012.

Flasar, F.M., V.G. Kunde, M.M. Abbas, R.K. Achterberg, P. Ade, A. Barucci, B. Bézard, G.L. Bjoraker, J.C. Brasunas, S.B. Calcutt, R. Carlson, C.J. Césarsky, B.J. Conrath, A. Coradini, R. Courtin, A. Coustenis, S. Edberg, S. Edgington, C. Ferrari, T. Fouchet, D. Gautier, P.J. Gierasch, K. Grossman, P. Irwin, D.E. Jennings, E. Lellouch, A.A. Mamoutkine, A. Marten, J.P. Meyer, C.A. Nixon, G.S. Orton, T.C. Owen, J.C. Pearl, R. Prangé, F. Raulin, P.L. Read, P.N. Romani, R.E. Samuelson, M.E. Segura, M.R. Showalter, A.A. Simon-Miller, M.D. Smith, J.R. Spencer, L.J. Spilker and F.W. Taylor. Exploring the Saturn System in the Thermal Infrared: The Composite Infrared Spectrometer, *Space Science Review 115,* 169-297, doi: 10.1007/s11214-004-1454-9, 2004.

Hedman, M.M., P. Helfenstein, R.O. Chancia, P. Thomas, E. Roussos, C. Paranicas, and A.J. Verbiscer, Photometric analyses of Saturn's small moons: Aegaeon, Methone and Pallene are dark; Helene and Calypso are bright. The Astronomical Journal 159, 129, 2020.

Howett, C.J.A., J.R. Spencer, J. Pearl and M. Segura, Thermal inertia and bolometric Bond albedo values for Mimas, Enceladus, Tethys, Dione, Rhea and Iapetus as derived from Cassini/CIRS measurements, Icarus 206, 573-593, 2010.

Howett, C.J.A., J.R. Spencer, T.A. Hurford, A. Verbiscer, M. Segura. Thermophysical property variations across Dione and Rhea. Icarus 241, 239–247, 2014.

Howett, C.J.A., J.R. Spencer, T.A. Hurford, A. Verbiscer, M. Segura. Maps of Tethys' thermophysical properties, Icarus 321, 705-714, 2019.

Howett, C.J.A., J.R. Spencer, T.A. Nordheim, Bolometric bond albedo and thermal inertia maps of Mimas, Icarus 348, 113745, 2020.

Nelder, J.A. and Mead, R. (1965). A Simplex Method for Function Minimization, *Computer Journal 7*, 308-313.

Nixon, C.A., T.A. Teanby, S.B. Calcutt, S.Aslam, D.E. Jennings, V.G. Kunde, F.M. Flasar, P.G.J. Irwin, F.W. Taylor, D.A. Glenar and M.D. Smith, Infrared limb sounding of Titan with the Cassini Composite InfraRed Spectrometer: effects of the mid-IR detector spatial responses, Applied Optics 48, 1912-1925, 2009

Nixon, C.A., M.S. Kaelberer, N. Gorius and the CIRS instrument team, User Guide to the PDS Dataset for the Cassini Composite Infrared Spectrometer (CIRS), GSFC STI/DAA Release Authorization #20120809115444, from the PDS website: https://pds-rings.seti.org/cassini/cirs/CIRS-User-Guide_v11.6-5-3-17.pdf, 2012

Pitman, K.M., Buratti, B.J., Mosher, J.A., Disk-integrated bolometric Bond albedos and rotational light curves of saturnian satellites from Cassini Visual and Infrared Mapping Spectrometer. Icarus 206, 537–560, 2010.

Royer et al. (2020) Dione and Helene Surfaces' History and Co-evolution, Icarus, in preparation.

Spencer, J.R. (1989) A rough-surface thermophysical model for airless planets, *Icarus 83*, 27-38, doi: 10.1016/0019-1035(90)90004-S.

Thomas, P., Tiscareno, M. S., Helfenstein, P. The inner small satellites of Saturn, and Hyperion. In: Schenk, P., Clark, R., Howett, C. J. A., Verbiscer, A., Waite, J. H., Dotson,

R. (Eds.), Enceladus and the Icy Moons of Saturn, pp. 387-408. University of Arizona Press, 2018.

Thomas, P. and P. Helfenstein, The small inner satellites of Saturn: Shapes, structures and some implication, Icarus In Press, 2019.

Umurhan O. M., A.D. Howard, J.M. Moore, P.M. Schenk and O.L. White. Reconstructing Helene's surface history, plastics and snow. In Lunar and Planetary Science XLVI, Abstract #2400. Lunar and Planetary Institute, Houston, 2015.

Verbiscer A., R. French, M. Showalter and P. Helfenstein. Enceladus: Cosmic graffiti artist caught in the act. Science, 315, 815, 2007.

Verbiscer, A., P. Helfenstein, B. Buratti, E. Royer. Surface properties of Saturn's moons from optical remonte sensing. In: Schenk, P., Clark, R., Howett, C. J. A., Verbiscer, A., Waite, J. H., Dotson, R. (Eds.), Enceladus and the Icy Moons of Saturn, pp. 323-341. University of Arizona Press, 2018.